\documentclass{elsart}
\begin{document}
\runauthor{Cugliandolo and Kurchan}
\begin{frontmatter}
\title{ Thermal properties of slow dynamics
\thanksref{X}}
\author[Paris]{Leticia Cugliandolo\thanksref{Someone}}
\author[Lyon]{\underline{Jorge Kurchan}\thanksref{Someone1}}
\thanks[X]{Invited talk to be presented at STATPHYS 20, Paris. \\ Paper dedicated to Heinz Horner on occasion
of his $60^{th}$ birthday.}
\address[Paris]{\it Laboratoire de Physique Th\'eorique de l'\'Ecole Normale 
Sup\'erieure,
\\
24 rue Lhomond, 75231 Paris Cedex 05, France \\ and \\
 Laboratoire de Physique Th\'eorique  et Hautes Energies, Jussieu\\
 5 \`eme \'etage,  Tour 24, 4 Place Jussieu, 75005 Paris France}
\address[Lyon]{{\it Laboratoire de Physique Th\'eorique de l' Ecole Normale Sup\'erieure de Lyon
 \\
46 All\'ee d'Italie, Lyon, France}}

\thanks[Someone]{leticia@physique.ens.fr}
\thanks[Someone1]{jkurchan@enslapp.ens-lyon.fr}

\begin{abstract}
 The limit of small entropy production is reached in relaxing systems
long after preparation, and in stationary driven systems in the limit of 
small
driving power. 
 Surprisingly, for extended systems this limit  is {\em not}
  in general the
 Gibbs-Boltzmann distribution,
or a small departure from it. Interesting cases in which it is not are
 glasses,  phase-separation,  
and certain driven complex fluids.
 
We describe a scenario with several
coexisting temperatures acting on different timescales, and partial 
equilibrations at each time scale.
This scenario entails strong modifications  of the fluctuation-dissipation 
equalities  and the existence  of 
some unexpected reciprocity relations. Both predictions are open
to experimental verification,
particularly the latter.
 
The construction  is consistent in general,  since it can be viewed  as the breaking
of a symmetry down to a residual group.
It does not assume the presence of quenched disorder.
It  can  be --- and to a certain extent has been ---  tested numerically,  while some experiments 
are on their way. There is  furthermore the   perspective that 
 analytic arguments may be constructed  to prove or
disprove its generality.

\end{abstract}

\begin{keyword}
Nonequilibrium extensive systems.

\noindent
{\em Pacs numbers:} 75.40.Gb, 75.10.Nr, 02.50.-r, 05.20.-y

\end{keyword}
\end{frontmatter}

\typeout{SET RUN AUTHOR to \@runauthor}

\section{Slow dynamics} 

\subsection{Two complementary aspects: driven and aging systems.}

  A universal feature of systems with slow dynamics is the
 extreme sensitivity of  their time-dependencies
to perturbations. Glassy systems  which `age' if left to relax 
(the  correlations evolve slower and slower as   time passes) may become stationary 
if an even  small amount of power  is pumped into them.
Thus, for example, a 3D spin-glass having the characteristic aging autocorrelation curves
(Fig. \ref{tres} $a)$ ), becomes stationary upon applying a very weak, 
slowly evolving
random field (Fig. \ref{tres} $c)$ ).
The reason  for this sensitivity is the presence of flat directions in phase-space, indeed
the very origin of the slowness of the dynamics.

Remarkably, some properties remain unaltered
even when the time-dependence has been dramatically changed by a small perturbation.
Precisely these properties  are  universal in slow  systems, at least qualitatively.

What this suggests is that we treat on an equal footing all problems  with slow dynamics,
 whether they age or are stationary \cite{Hoho}.  Their relevant common feature \cite{Cudeku}
(indeed, the definition of `slow') 
 is the smallness of the quantity that can be identified 
as an entropy production \cite{Degroot}
in each case. Thus, we  refer collectively as the `small entropy production' (SEP) limit
to the situation reached at long times and/or small external power input.   

\subsection{Fluctuation-dissipation temperatures.} 

In the SEP limit many systems are close to equilibrium. Such  situation has been well studied in the
past (see e.g. \cite{Degroot}). However, this is not the generic situation, 
particularly in complex systems 
such as glasses, phase-separating fluids, moving objects that are just unpinning, etc. 
An example will at this point clarify things. Consider a system with variables $\phi_i(t)$,
their correlation functions 
$ C_{ij}(t,t')  =\langle \phi_i(t) \phi_j(t') \rangle 
$,
 and the linear response $
R_{ij}(t,t') = 
\left. \frac{\delta \langle \phi_i(t) \rangle }{\delta h_j(t')}
\right|_{h=0} $ of  $\phi_i$ to a kick applied to $\phi_j$.
If we make a parametric plot of the integrated response 
\begin{equation}
\chi_{ij}( t+t_w,t_w)     =\int_{t_w}^{t+t_w} dt' \; R_{ij}(t+t_w,t')
\end{equation}
versus the corresponding correlation $ C_{ij}( t+t_w,t_w)$, the fluctuation-dissipation theorem 
is the statement  that the plot gives a straight line of gradient $-\beta=-1/T$. 
For an aging system we get instead, for large $t_w$, a limit curve that looks 
qualitatively like Fig. \ref{fig1}.
\begin{figure*}
\vspace{.3cm}
\centerline{\input{chiCeps0T02.pslatex}}
\vspace{.3cm}
\caption{Integrated response versus correlation plot for a mean-field glass model.}
\label{fig1}
\end{figure*}
This figure has been obtained for the autocorrelation of the variables in a 
mean-field glass model \cite{review},
but similar plots have been obtained numerically for 3 and 4 dimensional spin-glasses \cite{Roma1},
Lennard-Jones glasses with Montecarlo \cite{Parisi} and molecular dynamics
\cite{Koba}. Ordinary domain growth  (as well as droplet models
for spin-glasses)   \cite{Abarrat} also yield such a curve, with the peculiarity
that the line to the left becomes {\em horizontal}. 

 These systems  age. As mentioned above,  we can 
 perturb them with small,
nonconservative forces which render  them stationary. 
Remarkably, at least
in every model we know, if  we now make the plots of Fig. \ref{fig1}
we get, even for small driving powers,
{\em the same dependence} \cite{Ho,Cukulepe}.  
Despite the fact that the driven systems are stationary,
they are well out of equilibrium.

Such robustness of the plot in Fig. \ref{fig1} suggests that the fluctuation-dissipation ratio
 might have a  physical meaning. Let us define, for any two observables
and for any two times, the effective temperatures as \cite{Cukupe}
\begin{equation}
\beta^{ eff}_{ij}(\omega,t_w)
\equiv
\frac{\chi_{ij}''(\omega,t_w)}{\omega {\mbox{Re}} \tilde C_{ij}(\omega,t_w)} 
\label{defi1}
\end{equation}
We have Fourier transformed the time-difference, as in experiments. In a stationary system the
$t_w$ dependence dissappears.

\subsection{Scales}

Equation  (\ref{defi1}) is just a definition.
The scenario  we shall discuss  here consists of saying that 
{\em for all pairs of observables and for each timescale there is a single effective temperature}.
That is,
\begin{equation}
 \!\!\!\!\!\!\!\!\!
 \beta_{ij}^{ eff}(\omega,t_w) - \beta_{kl}^{ eff}(\omega,t_w) \sim 0  \;\;\;\;\;\;  
{\mbox{or}} 
\;\;\;\;\;\;
\beta_{ij}^{ eff}(\omega,t_w) \sim \beta^{ eff}(\omega,t_w) \; ,\;\;\; \forall \; i,j
\label{terma}
\end{equation}
and
\begin{equation}
\beta^{ eff}(\omega_1,t_w)-\beta^{ eff}(\omega_2,t_w) 
\sim 0 \;\;\;\;\;\;\;\;\;\;\;\; {\mbox{if}} \;\; 
\frac{\omega_1}{\omega_2}={\mbox{const.}}
\label{scala}
\end{equation}
in the SEP limit ($t_w \rightarrow \infty$ and vanishing  driving power).
A system with a  $\chi$ {\em vs.} $C$ plot like  Fig. \ref{fig1} (two straight lines) then has two
 temperatures (one of them being $T$). In 
systems with more than two
temperatures the plot has a straight line of gradient $-1/T$ plus a piece of non-constant slope. 

\begin{figure}[h]
\vspace{.3cm}
\centerline{\input{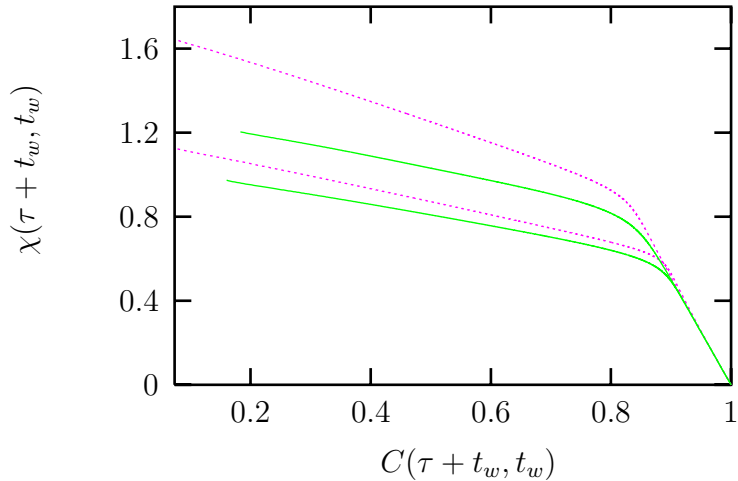}}
\vspace{.3cm}
\caption{Thermalisation of two subsystems. Integrated response vs. correlation plots
for two  uncoupled  systems (dots) and for systems that are coupled (lines).}
\label{fig2}
\end{figure}

\subsection{Thermometry. Zeroth law.}

One can now ask if (\ref{defi1}) really defines temperatures.
To check this,  one couples a `thermometer' system simultaneously
to the same  observable  of an
ensemble of many independent copies of the system \cite{Cukupe,long}. One can then
show that the copies of the system act on the thermometer as a superposition  of thermal baths
with the corresponding effective temperatures and timescales.
In order to measure a single temperature, one can construct a tuned thermometer (e.g. an oscillator) 
that responds 
to a single time-scale.  

To see the relation of FDT-temperatures  with the zeroth law, one can couple  the `thermometer'
as before, but alternatively to two observables. Then one shows that there is on average a net
heat flow from the higher to the lower effective temperature.

\subsection{Thermalisation of subsystems}

Within the present scenario, two systems with different effective temperatures
if weakly coupled eventually `thermalise' as follows: If the coupling is relatively
 strong \cite{Cukupe}, 
we have the situation of Fig. \ref{fig2} (a direct solution of the equations, without assumptions):
 two (mean-field) glassy systems were defined so as to 
have different effective temperatures when uncoupled 
(dotted lines). If they are made to evolve coupled,
the  effective temperatures equalise (full lines).

If the coupling is weak
  \cite{long}, the system 
preserves essentially the temperatures  of its constituents, but eventually
rearranges them 
in widely separated timescales. Thus, the combined system has three temperatures, 
the bath temperature plus other two, each acting on a separate timescale.

\subsection{Auxiliary thermal baths}

\begin{figure*}
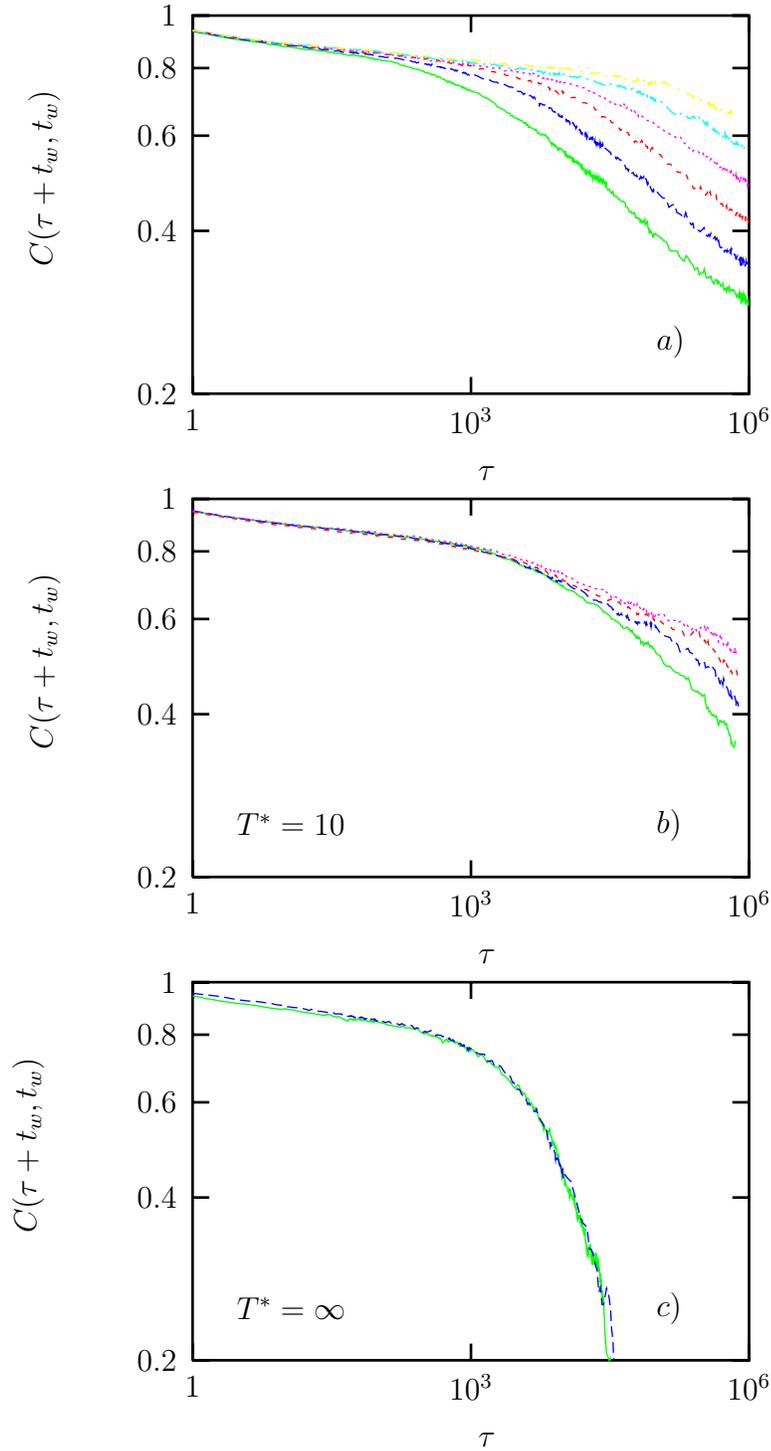

\centerline{\input{cT1.4color.pslatex}}
\vspace{.3cm}
\centerline{\input{T1.4Ts10nu01g5000color.pslatex}}
\vspace{.3cm}
\centerline{\input{T1.4bs0nu01g10000color.pslatex}}
\vspace{.3cm}
\caption{Autocorrelations of 3D Edwards-Anderson model.
From top to bottom: a) pure relaxtion, b) same but coupled to a weak slow bath with $T^*=10$, c)
  same but coupled to a weak slowly evolving random field (a weak slow bath with $T^*=\infty$).
In b) aging has partially dissappeared, while in c) it has dissappeared completely.
}
\label{tres}
\end{figure*}

An interesting tool for the study of a system with more than one `natural' temperature
is to couple it to a {\em slow, weak} auxiliary bath \cite{long} of 
temperature $T^*$, which, as we
shall see, plays the role of  
 `a field conjugate to the natural temperatures' \cite{rem}.
The outcome, within this scenario,  is that the auxiliary bath fixes the timescale
at which its temperature happens in the system. In particular, 
an aging system with a single $T^{ eff}>T$ will become 
stationary if $T^*>T_{ eff}$, and be hardly affected if $T^*<T^{ eff}$. More generally, 
an aging  system with multiple effective temperatures should  become partially stationary (for  
all the timescales with $T^*>T^{ eff}$), but would still have aging for timescales with 
$T^*<T^{ eff}$. 
Simulations for the three dimensional Edwards-Anderson model 
(a system which seems to have many
temperatures \cite{Roma1})
 are in this sense very encouraging
(see Fig. \ref{tres}).

\subsection{Reciprocity}

A surprising feature of the present scenario  is that 
in the  SEP limit there are 
 reciprocity relations 
$
\langle A(t) B(t') \rangle = 
\langle B(t) A(t') \rangle
$, 
and similarly   
for the responses.

It is remarkable that reciprocity relations 
 hold in a situation in which FDT is strongly violated. 
This is all the more surprising in an
aging case in which the system is not even stationary. 
The interest of these relations is that they are relatively easy to measure
in a simulation or in an experiment.  

\begin{figure}[h]
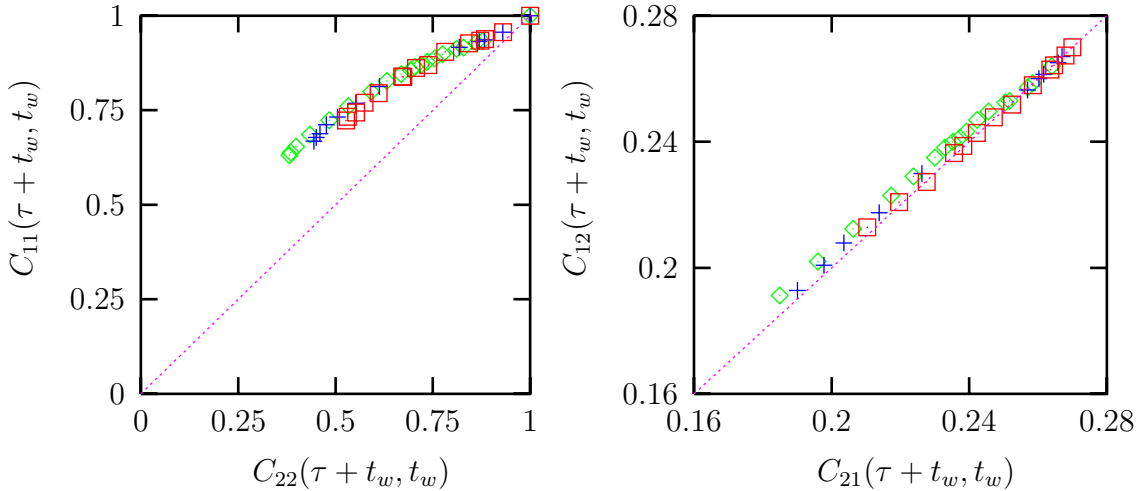

\vspace{.3cm}
\centerline{\input{3DEAonsaautocolor.pslatex}\hspace{-1cm}\input{3DEAonsacrosscolor.pslatex}}
\vspace{.1cm}
\caption{Parametric plots of $C_{11}$ {\em vs.} $C_{22}$ (left) and $C_{12}$ {\em vs.} $C_{21}$ (right)
 for $t_w=3000$ (diamonds), $t_w=10000$ (crosses), $t_w=30000$ (squares).
Note that the points in the right graph approach the diagonal for larger $t_w$. }
\label{fig7}
\end{figure}

Figure \ref{fig7} shows a numerical check for two 3D spin glasses. $C_{11}$ and $C_{22}$
denote the autocorrelations of systems 1 and 2 (which have coupling independent constants, 
on average  half as strong in one system than  in the other).  Despite the assymmetry between the two
systems (cfr. the left figure), the mutual correlations  $C_{12}$ and   $C_{21}$ tend to equalise
more and more for longer waiting times (right figure).

\subsection{Emergence of macroscopic temperatures.}

A very encouraging feature of the effective temperatures, such as they come out in 
the solution of some analytic models \cite{review}, is that they remain 
non-zero in the limit of zero
bath temperature.
In other words, {\em the temperatures stay non-zero even when  Boltmann's constant is negligible}.
This is  most welcome: if these effective temperatures
are to be relevant for macroscopic cases such as colloids, sheared foams and (perhaps) granular media,
they must be huge in units of the Boltmann constant, and be related to macroscopic structures.

\section{Analytical results. Symmetry breaking}

Let us outline very briefly the analytical arguments 
showing that this scenario `closes'.
We assume the system has Langevin dynamics:
\begin{equation}
-m_i {\ddot \phi}_i -  M_{ij} \phi_j - \frac{
\partial V({\bf \phi})}{\partial \phi_i}   =
 \Gamma_0 {\dot \phi}_i - \eta_i
\label{motion}
\end{equation}
where $\eta_i$ are uncorrelated Gaussian white noises with variance $2\Gamma_0 T$
and $\Gamma_0$ is the strength of the coupling to the `white'  bath.
One can also add weak nonconservative forces.
The masses $m_i$ can in particular  be zero. 

We can encode the correlations and (causal) responses in the `superspace' order parameter
(see {\em e.g.} \cite{Modecoup}):
\begin{equation}
 Q_{ij}(1,2)
=
C_{ij}(t_1,t_2) +
(\bar \theta_2 - \bar \theta_1) \;
\left[
\theta_2 \,
\; R_{ij}(t_1,t_2)
- \theta_1 \, \; R_{ji}(t_2,t_1)
\right]
\; .
\label{Q12}
\end{equation}
where $\theta_a$, $\bar \theta_a$ are Grassmann variables, and we denote the full set of coordinates 
in a  compact form as
$1= t_1 \theta_1 \overline\theta_1$, $d1=   dt d\theta d\overline\theta$, etc.

One can obtain a general perturbative expression 
for the correlations and responses (\cite{long}, see also 
\cite{Modecoup}):
\begin{equation}
 \! \! \! \!\! \! \! \!\! \! \! \!\! \! \! \!\! \! \!
\left( m_i \frac{\partial^2}{\partial t_1^2} + 
       \Gamma_0 D_1^{(2)} \right) \, Q_{ij}(1,2) + 
       M_{ik}  \, Q_{kj}(1,2) 
= 
\delta(1-2)\delta_{ij} +
 [{\bf \Sigma} \otimes {\bf Q}]_{ij}(1,2) 
\label{motionc}
\end{equation}
where $\otimes$ stands for convolution and matrix product.
  The function
${\bf \Sigma}$ is obtained from all diagrams with propagator ${\bf Q}$
with two (amputated) legs that cannot be disconnected by cutting one 
more line.
{\em Note that we have nowhere assumed there is quenched disorder, and that this
expansion can be done for any model, independently of the dimensionality}.
 
We now  make a separation between `fast' ${\bf Q^F}$ and `slow' ${\bf {\tilde Q}}$
evolution $
{\bf Q}={\bf Q^F} + {\bf {\tilde Q}}$. 
In the `fast' 
regime of finite time-differences, correlations and responses can be shown 
\cite{Cudeku} to satisfy  FDT, and
are time-translation invariant. In the SEP limit one can obtain  a separate
equation for the `slow' part:
\begin{equation}
 \!\!\!\!\!\! M_{ik} {\tilde Q}_{kj}(1,2) -
 \frac{    q^d_{kj} -q_{kj}^{EA}    } {T} {\tilde \Sigma}[\tilde Q]_{ik} (1,2)
- \frac{    d^d_{kj} -d_{kj}^{EA}    } {T} {\tilde Q}_{ik} (1,2)  = {\mbox{`small'}}
\label{motionsl}
\end{equation}
Here ${\mbox{`small'}}$ stands for terms that vanish in the SEP limit (see below), and
${\tilde \Sigma}[\tilde Q]_{ik} (1,2)$ is a functional of $\tilde Q$ that is also
 obtained diagrammatically. The $q^d_{kj},q_{kj}^{EA},d^d_{kj},d_{kj}^{EA}$ are 
parameters to be determined
from (\ref{motionsl}) and the equations for ${\bf Q^F}$. 
 
\subsection{Symmetries}

In general,  to the extent that one is 
allowed to neglect the `small' terms in the SEP limit, 
Eqs.  (\ref{motionc}) are invariant with respect to any change of `coordinates'
$t_a$, $\theta_a$, and ${\bar \theta}_a$ ($a=1,2,...$) with 
unit superjacobian \cite{Frku}.
This is a large symmetry group, including  in particular the following:

\begin{itemize}

\item[i.] {\it Ordinary time-reparametrizations.}

\begin{equation}
t_a \rightarrow K(t_a) 
\; , \;\;\;\;\;\;\;\;\;\;
\theta_a \rightarrow {\dot K}(t_a) \theta_a
\; , \;\;\;\;\;\;\;\;\;\;  
{\bar \theta}_a \rightarrow {\bar \theta}_a
\; .
\label{repp}
\end{equation}
 These reparametrizations 
were already considered
by Sompolinsky and Zippelius in their seminal work \cite{So}.
Under this subgroup the order parameter ${\bf {\tilde Q}}$
preserves the form (\ref{Q12}).

\item[ii.] {\it Time-reversal.}

\begin{eqnarray}
t_a \rightarrow -t_a - \beta^* \overline\theta \theta 
\; , \;\;\;\;\;\;\;\;\;\;
\theta_a \rightarrow  {\bar \theta}_a 
\; , \;\;\;\;\;\;\;\;\;\;  
{\bar \theta}_a \rightarrow  \theta_a
\; .
\label{repp1}
\end{eqnarray}
These reparametrizations only preserve the form (\ref{Q12}) if
they act on a  ${\bf {\tilde Q}}$ such that
$
R_{ij}(t-t') 
= 
\beta^* \partial_{t'} C_{ij}(t-t') \, \Theta(t-t')
$.

\item[iii.] {\it Boson-Fermion symmetry }

A group  of symmetry transformations which, unlike the previous two,
exchanges bosons and fermions is
\begin{equation}
 \! \! \! \!\! \! \! \!\! \! \! \!\! \! \! \!\! \! \! \!\! \! \! 
 K(t_a) \rightarrow K(t_a) + K_0
 \; ,\;\;
{\bar \theta}_a    \rightarrow      {\bar \theta}_a  + {\bar\epsilon}    
 \; , \;\;
{ \theta}_a  \rightarrow 
{ \theta}_a +  { \tilde \epsilon}    \;, \;\;
t_a \rightarrow  t_a +\frac{ \epsilon}{{\dot K}(t)} {\bar \theta}_a 
\label{14j}
\end{equation} 
for any $K(t)$, $K_0$, ${\bar\epsilon}$, ${ \tilde \epsilon} $, $\epsilon$ (the latter 
three, odd-Grassmann
parameters).

\end{itemize}

\subsection{Temperature fixing as symmetry breaking}

Let us now briefly show how this symmetry considerations 
prove the consistency of the ansatz. We shall for brevity
only mention here the case of a single 
effective temperature $T^{eff}$ for the slow dynamics.

Equation (\ref{motionsl}) (neglecting its r.h.s.)  will always admit a solution
invariant with respect to  some residual subgroup  of the group
of transformations with unit superjacobian.   
To get a non-constant solution, we must then break (i), by choosing a
 time-reparametrisation  $K(t)$. One can easily check
that fixing an effective temperature now corresponds to
 breaking (iii) to a three-parameter subgroup with fixed $K(t)$ and with the restriction
${\tilde \epsilon}=T^{eff} \epsilon$ in  (\ref{14j}).
One can further keep `time reversal' (\ref{repp1}), but as applied {\em on the
reparametrised time} $K(t)$, and with $T^*=T^{eff}$. 

This may seem rather strange, but in fact it is the abstract way of describing 
 the usual temperature fixing by a thermal bath.
{\em In any purely Hamiltonian system} ($\Gamma_0=0$), 
Eqs.~(\ref{motionc}) are invariant under the  group (\ref{14j}) 
with $K(t)=t$ and four parameters
$K_0$, $\epsilon$, ${\tilde \epsilon}$ and ${\bar \epsilon}$.
If we now switch on a thermal bath of temperature $T$ (we make $\Gamma_0>0$), 
the group (\ref{14j}) is broken down by the `bath' term proportional to
$\Gamma_0$ to the residual  three parameter subgroup \cite{Gozzi} obtained by restricting
 ${\tilde \epsilon} = T \epsilon$.  

\subsection{Sensitivity and the matching problem}

The neglect of the `small' terms in the right hand side of
equation (\ref{motionsl}) brings about 
the invariance with respect to super-reparametrisations.
However, the true equations do not have so many  symmetries.
In order to see what is the situation, we note that
equation (\ref{motionsl}) is of the form:
\begin{equation}
\left[ \mbox{ (Super) Reparametrization Invariant } \right] = 
\mbox{ `small', non-invariant }
\end{equation}
If we find one solution to this equation by 
setting the l.h.s.  to zero, 
any super-reparametrization is also a solution.
If the solution to l.h.s.=0  
breaks reparametrization invariance, one  obtains thus an infinite number  
of solutions.
In order to fix the correct one, 
one must take into account the `small' terms in the r.h.s.
This requires a  calculation going beyond  the zero
entropy production limit (small but finite forcing or long but finite waiting times), and
this is a much harder task, indeed a problem that remains to be solved.

On the other hand, apart from being a technical nuisance, 
this has a physical meaning. 
Because  terms that are arbitrarily small
 decide which reparametrization is the good one,
 systems having non-trivial solutions to (\ref{motionsl})
 {\em  are sensitive to
 perturbations that are arbitrarily weak}, a general feature
of systems with slow dynamics.

\section{Perspectives}

Given the number of not entirely obvious relations this scenario  proposes,
the least one can now say is that 
it  has  heuristic value, even if the fact of having been 
inspired by mean-field spin-glass theory
is sometimes
viewed as an original sin.
There has been recently quite a lot of numerical activity 
to test the predictions,  with encouraging results. 
The experimental exploration of  structural glasses
 from this point of view
is also now beginning.

The solution is  consistent in general, since it 
corresponds to
 the breaking of a  symmetry  that is a general property of  slow evolution down to a 
residual subgroup
(having the physical meaning of partial thermalisations). 
Just as in any partial symmetry breaking situation, one cannot with such arguments alone 
 assure that in some model this symmetry will not   be broken further, or in a different way.    
One may perhaps hope that real proofs concerning the generic situation 
in the SEP limit of extended systems will be one day available.


\begin{thebibliography}{999}


\bibitem{Hoho} This point of view has been advocated by Horner \cite{Ho}. See also
\cite{Cukulepe}.

\bibitem{Ho} H. Horner, {\em Z. Phys.} {\bf B57}
  (1984) 29; {\it ibid.} (1984) 39.

\bibitem{Cukulepe}
L. F. Cugliandolo, J. Kurchan, P. Le Doussal and L. Peliti; 
{\em Phys. Rev. Lett.} {\bf 78} (1997), 350.


\bibitem{Cudeku}
L. F. Cugliandolo, D.S. Dean and  J. Kurchan;
{\it  Phys. Rev. Lett.} {\bf 79} (1997)  2168.

\bibitem{Degroot}
R. Kubo, M. Toda and N. Hashitume;
{\it Statistical Physics II. 
Nonequilibrium Statistical Mechanics}, Springer-Verlag, 1992; \\
S. R. De Groot and P. Mazur; {\it Non-equilibrium thermodynamics},
Dover Pub., New York, 1984. 


\bibitem{Roma1} E. Marinari, G. Parisi,  
F. Ricci-Tersenghi and J.J. Ruiz-Lorenzo;
 {\em J. Phys.} {\bf A 31} (1998) 2611.

\bibitem{Parisi} G. Parisi,
{\it  Phys. Rev. Lett.} {\bf 79} (1997) 3660.

\bibitem{Koba} J-L Barrat and  W. Kob, cond-mat/9806027. 

\bibitem{Abarrat} A. Barrat, {\it Phys. Rev.} {\bf E57} (1998) 3629. 


\bibitem{Cukupe}
L. F. Cugliandolo, J. Kurchan and L. Peliti;  
{\em  Phys. Rev.} {\bf E55} (1997) 3898.

\bibitem{long} L.F. Cugliandolo and J. Kurchan,
{\em `Thermal properties of systems with slow dynamics'},
in preparation.

\bibitem{rem} This is much in the spirit of: 
G. Parisi and M.A. Virasoro {\em J. Physique} (Paris) {\bf 50} (1986) 3317, and 
R. Monasson, {\it Phys. Rev. Lett.} {\bf 75} 2847 (1995).

\bibitem{review}  
 J-P Bouchaud, L. F. Cugliandolo, J. Kurchan and M. M\'ezard;
cond-mat/9702070; 
 {\em Spin-glasses and random fields}, A. P. Young ed.
 (World Scientific, Singapore 1998).


\bibitem{Modecoup}J.P. Bouchaud, L.F. Cugliandolo,  J. Kurchan et M. M\'ezard. 
{\em Physica}  {\bf A226} (1996) 243.


\bibitem{Frku} S. Franz and J. Kurchan. 
{\em Europhys. Lett.}  {\bf 20} (1992) 197.

\bibitem{So} H. Sompolinsky, {\em Phys. Rev. Lett.} {\bf 47} (1981)  935;
 H. Sompolinsky and A. Zippelius, {\em Phys. Rev.} {\bf B25}
(1982)  6860.


\bibitem{Gozzi} E. Gozzi; {\em Phys.Rev.} {\bf D30} (1984) 1218.

\end{thebibliography}
\end{document}